\documentclass[12pt,letterpaper,amsmath, showpacs, amssymb]{iopart}

\usepackage{graphicx}
\usepackage{hyperref}

\def\longrightharpoonup{\relbar\joinrel\rightharpoonup}
\def\longleftharpoondown{\leftharpoondown\joinrel\relbar}

\def\longrightleftharpoons{
  \mathop{
    \vcenter{
      \hbox{
	\ooalign{
	  \raise1pt\hbox{$\longrightharpoonup\joinrel$}\crcr
	  \lower1pt\hbox{$\longleftharpoondown\joinrel$}
	}
      }
    }
  }
}

\newcommand{\rates}[2]{\displaystyle
  \mathrel{\longrightleftharpoons^{#1\mathstrut}_{#2}}}

\begin{document} 

\title{On the formation and decay of a molecular ultracold plasma}
\author{N. Saquet, J. P. Morrison, M. Schulz-Weiling, H. Sadeghi, J. Yiu, C. J. Rennick and E. R.  
Grant}
\address{Department of Chemistry, University of
British Columbia, Vancouver, BC V6T 1Z3, Canada}

\date{\today}

\begin{abstract}

Double-resonant photoexcitation of nitric oxide in a molecular beam creates a 
dense ensemble of $50f(2)$ Rydberg states, which 
evolves to form a plasma of free electrons trapped in the potential well of an 
NO$^+$ spacecharge. The plasma travels at the velocity of the molecular beam, 
and, on passing through a grounded grid, yields an electron time-of-flight 
signal that gauges the plasma size and quantity of trapped electrons. This 
plasma expands at a rate that fits with an electron temperature as low as 5~K, colder that typically observed for atomic ultracold plasmas.  The recombination of molecular NO$^+$ cations with electrons forms neutral molecules excited by more than twice the energy of the NO chemical bond, and the question arises whether neutral fragmentation plays a role in shaping the redistribution of energy and particle density that directs the short-time evolution from Rydberg gas to plasma. To explore this question, we adapt a coupled rate-equations model established for atomic ultracold plasmas to describe the energy-grained avalanche of electron-Rydberg and electron-ion collisions in our system.  Adding channels of Rydberg predissociation and two-body, electron- cation dissociative recombination to the atomic formalism, we investigate the kinetics by which this relaxation distributes particle density and energy over Rydberg states, free electrons and neutral fragments.  The results of this investigation suggest some mechanisms by which molecular fragmentation channels can affect the state of the plasma.   

\end{abstract}

\pacs{52.55.Dy, 32.80.Ee, 33.80.Gj, 34.80.Lx}

\section{Introduction}

Simple two-body pictures define much of our understanding of the natural world.  From the properties of gases to the kinetics of reactions in solution, model systems of atoms and molecules evolve to observable effect by means of short-range interactions, or no intermolecular interactions at all.   The many-body physics of charged particle correlation confounds this picture in the case of ionized systems.  

Many body Coulomb interactions begin to govern the dynamics of charged particles in a plasma when the electrostatic potential energy exceeds the thermal energy in translational degrees of freedom.  The point at which this occurs depends on the density and temperature of the plasma \cite{Ichimaru}.  The parameter, $\Gamma$, gauges the degree of correlation in terms of the dimensionless ratio,   
\begin{equation}
	\Gamma  = \frac{e^{2 }/4\pi \varepsilon _{0} a}{kT},
\label{equ:correlation}
\end{equation}

\noindent where $e$ is the charge, and $a$ is the Wigner-Seitz radius, which relates to the particle density, $\rho$, by, 

\begin{equation}
	4/3~\pi a^3  = 1/\rho.
\label{equ:wigner}
\end{equation}

\noindent The effects of correlation are complex and difficult to study, but play an important role in shaping behaviour on a scale as large as stars and as small as quantum dots \cite{Gabadadze,Nelson}.  Ultracold plasmas offer one of the simplest possible means to approach the conditions of strong Coulomb coupling.   

Well-established methods exist to form ultracold plasmas starting with laser-cooled atoms in magneto-optical traps (MOT) \cite{Gallagher2003,Killian_Phys_rept}.  The many experimental studies of such systems have provided a revealing glimpse of ionized gas dynamics under conditions that approach those necessary for ion and electron correlation.  Important conclusions from such studies include the following:  Photo-prepared atomic systems in MOTs evolve to plasma when excited both to levels above the ionization threshold and to Rydberg states just below.   Disorder-induced heating elevates ion temperatures to near 1 K.  Three-body recombination, followed by electron-Rydberg atom super-elastic collisions heat free electrons to temperatures of 30 K or more, as measured by plasma expansion.  At typical MOT densities below $10^{10}$ cm$^{-3}$, T$_e$ values this high yield correlation parameters, $\Gamma_e$, less than 1.  

Seeded supersonic expansions produce moving-frame temperatures in the range of 1 K or less.  Typical densities of active species in laser-crossed-beam interaction volumes can substantially exceed those attainable in MOTs, resulting in much more concentrated charged particle distributions.  Unlike optical atom traps, free-jet expansions can easily incorporate molecules.  

We have adopted a supersonic beam strategy to form an ultracold plasma of nitric oxide molecular cations and electrons \cite{Morrison2008,Morrison2009}.  In these experiments, we use double-resonant laser excitation of nitric oxide, cooled to 1 K in a seeded supersonic molecular beam, to produce a Rydberg gas of $\approx 10^{12 }$ molecules cm$^{-3}$ in a single selected state. This population evolves to produce prompt free electrons and a durable cold quasineutral plasma of electrons and intact NO$^{+}$ ions. Thereafter, a field of amplitude as low as 3 V/cm, far smaller than that required to field-ionize the precursor state, extracts a small signal of electrons, but pulses of amplitude as high as 200 V/cm, fail to destroy the plasma. These observations signal two properties that confirm the evolution to a cold plasma:  The presence of a surface charge of electrons, extractable by a weak field \cite{Haroche}, and a Debye screening length much smaller than the diameter of the plasma, shielding the charged particles in its core \cite{Rolston1999}. 

Downstream, transmission through a moveable grid samples the electrons in this core, profiling their spatial distribution as a function of time.  We find that the volume increases in accord with an electron-pressure-driven ambipolar expansion.  Analysis of this expansion in the simple limit of the Vlasov equations returns temperatures that appear lower than those observed for atomic plasmas in MOTs.  

This molecular plasma differs from atomic ones both by virtue of its density and the fact that its positively charged component consists of molecular NO$^+$ cations.  The recombination of these cations with electrons forms neutral molecules excited by more than twice the energy of the NO chemical bond.  Plasma dissipation to neutral atom fragments surely affects the long-time dynamics of this system.  It may as well play a role in shaping the redistribution of energy and particle density that directs the short-time evolution from Rydberg gas to plasma.  

In an effort to explore these questions, we have adapted a coupled rate-equations model established for atomic ultracold plasmas \cite{Pohl_atom} to describe the energy-grained avalanche of electron-Rydberg and electron-ion collisions in our system.  We model these kinetics as they distribute particle density and energy over Rydberg states and free electrons, driving the short-time evolution from Rydberg gas to plasma.  

Our experimental observations gauge plasma dynamics on a tens-of-microsecond timescale.  A fuller model calculation will describe processes that unfold over this entire time, accounting both for a spatial variation in the charged particle density and its ambipolar expansion.  For the present, we focus on the initial evolution to plasma in our molecular system by adding channels of Rydberg predissociation and two-body, electron-cation dissociative recombination to the atomic formalism.  Although preliminary in this respect, our results clearly demonstrate the extent to which predissociation can be expected to have a distinct, observable effect on the state of the plasma.  Here we present experimental measurements detailing the time variation of our plasma width and electron density, interpreted within the framework of calculations based on this preliminary model.   

\section{Experimental}

\begin{figure}[ht]
    \begin{center}
        \includegraphics{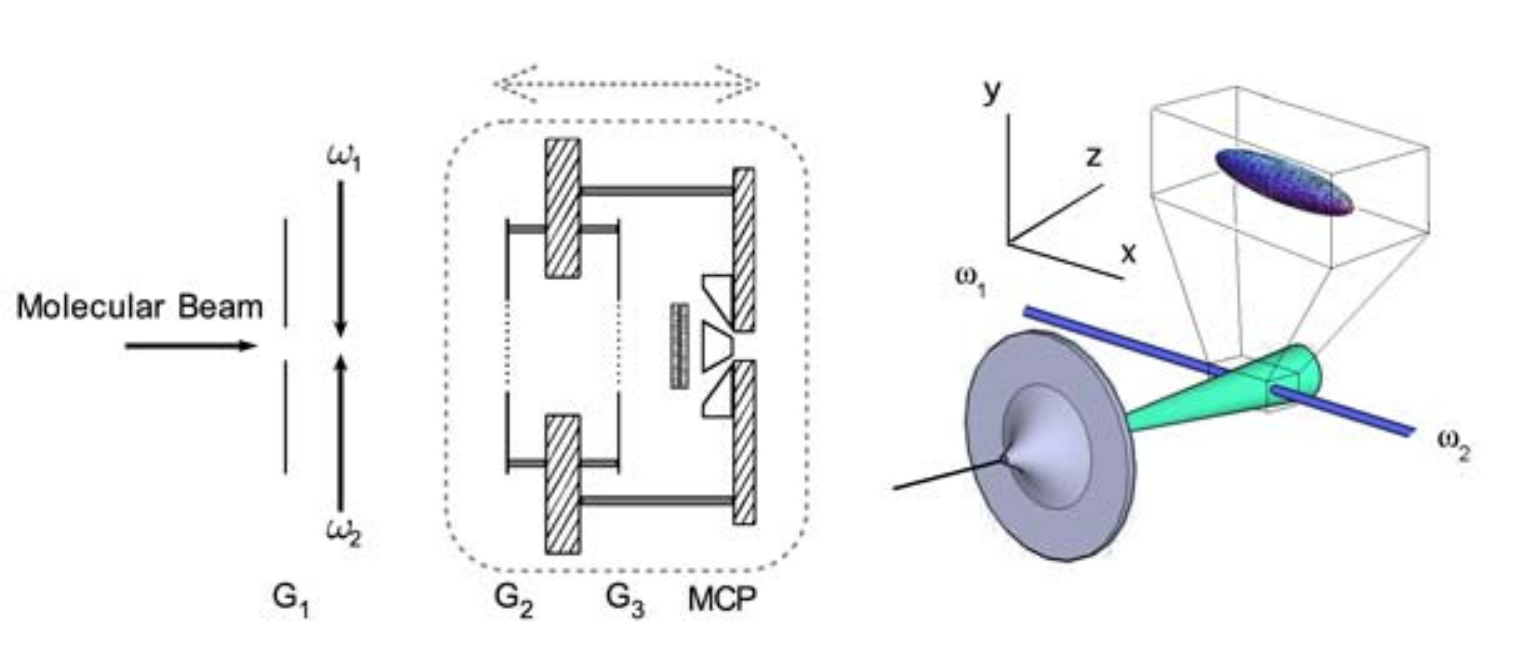}
    \end{center}
    \caption{Left:  Experimental apparatus. A skimmed molecular beam passes through an aperture in plate G$_1$, after which it intersects counter-propagating laser beams indicated as $\omega_1$ and $\omega_2$. As the excitation volume transits the plane defined by grid, G$_2$, a microchannel plate detector (MCP) situated behind grid, G$_3$, collects the signal of extracted plasma electrons.  The components surrounded by the light-grey border translate together on a moving carriage.  Right:  Image of the illuminated volume formed by the intersection of laser beams $\omega_1$ and $\omega_2$ with the molecular beam.}
    \label{fig:apparatus}
\end{figure}

Figure \ref{fig:apparatus} diagrams the apparatus.  A pulsed free jet expansion of NO seeded at 1:10 in He forms a differentially pumped molecular beam as it enters the main chamber through a 1 mm diameter skimmer.  There it passes through the first of three perpendicular plates into a field-free region where it crosses a pair of counter-propagating laser beams. The first dye laser pulse ($\omega_1$) excites ground state NO (X$^{2}\Pi_{1/2} ~v=0, J=1/2$) to the ground rovibronic level of the A~$^{2}\Sigma^+$ state. The second frequency-doubled dye laser ($\omega_2$), tuned over the range from 330 to 340~nm, promotes transitions from this intermediate A-state to selected Rydberg states of principal quantum number, $n$, from 35 to the ionization limit.

The nozzle diameter (0.5 mm) and total backing pressure (4 atm) of our beam source predict a centerline NO density of $3.6 \times 10^{13}$ cm$^{-3}$ at the point of illumination.  At a measured rotational temperature of 2.5 K, 89\% of these molecules populate the two parity components of the rotational ground state.  We adjust the power of $\omega_1$ and $\omega_2$ to saturate both steps of excitation, driving 50 percent of the population in the negative parity component of the ground state to the A state, and 50 percent of the A-state population to the selected high-Rydberg state. On this basis, we estimate the peak density of high-Rydberg NO* molecules available for plasma formation to be $5\times 10^{12}$ cm$^{-3}$.  

The present laser-crossed molecular beam excitation geometry produces a prolate ellipsoid excitation volume with an estimated aspect ratio of $\sim$4.8:1.  The width of the molecular beam determines the long axis of this ellipsoid. The short axes conforms with the two-dimensional Gaussian intensity distribution of the spatially filtered $\omega_1$ laser beam, collimated to a full-width at half-maximum of $758 ~\mu m$ \cite{Morrison2009}. The plasma produced in this volume travels with the velocity of the molecular beam over an adjustable distance to pass through the $(x,y)$ vertical plane of the imaging grid, G$_2$. 

The carriage supporting this grid, together with G$_3$ and the MCP detector, travels on linear bearings supported by three 0.5~inch diameter stainless steel rods. A bellows-isolated motorized actuator controls the position of this carriage to within 10~$\mu$m over a range of 10~cm. 

We set G$_2$ to the same potential as G$_1$ (nominally apparatus 
ground), creating a field-free region in which we vary the plasma 
time-of-flight. Upon transiting G$_2$, the plasma enters an electrostatic field 
determined by the potential applied to the third grid, G$_3$, spaced downstream 
by a fixed distance of 16~mm. For the present experiments, we apply 145~V to 
G$_3$. As the plasma traverses G$_2$, the field between G$_2$ and G$_3$ 
extracts electrons which, on the timescale of the measurement, appear 
instantaneously as signal at the multichannel plate detector. This signal 
represents the extracted electron density as a function of $z$, integrated in the transverse $(x,y)$ 
dimensions of the plasma, providing a time-dependent trace of this changing density as 
the plasma volume passes through G$_2$. We assume that the electrons extracted 
by G$_2$ supply a representative gauge of the short-axis plasma width and relative charge 
density. 

\section{Results}

In previous work, we have shown that a gas of high-Rydberg NO molecules, entrained in a molecular beam with a longitudinal temperature of \it ca.~\rm700~mK, evolves to form a quasineutral ultracold plasma of NO$^{+}$ ions and electrons  \cite{Morrison2008, Morrison2009}. Present experiments measure the expansion and decay of this plasma by tracking its electron density waveform as a function of time-of-flight over adjustable intervals of distance.
 
\begin{figure}[ht]
    \begin{center}
        \includegraphics[height=4 in]{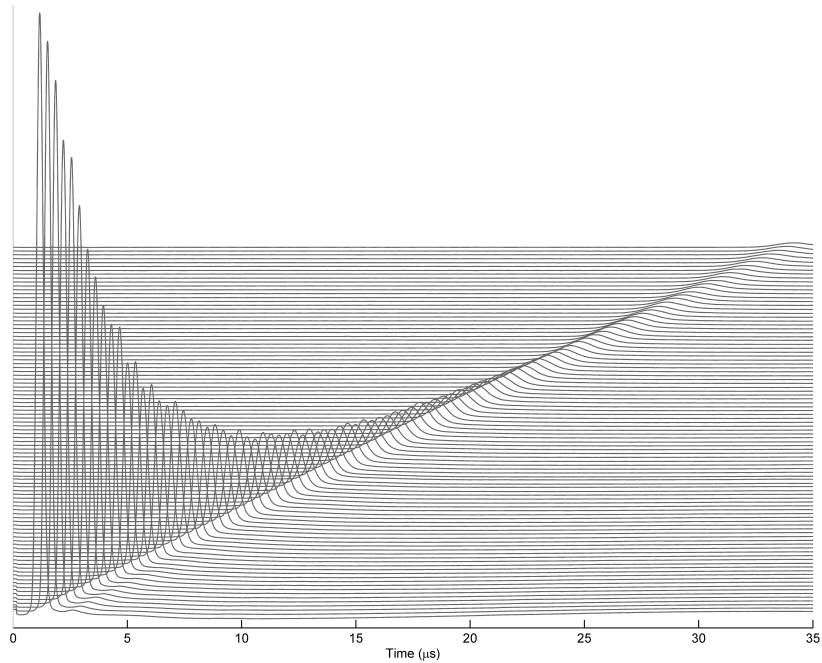}
    \end{center}
    \caption{Waveforms captured by the MCP detector for different positions of the 
    moving carriage. Traces have been offset vertically for clarity.}
    \label{fig:stackplot}
\end{figure}

\begin{figure}[ht]
    \begin{center}
      \includegraphics[height=4in]{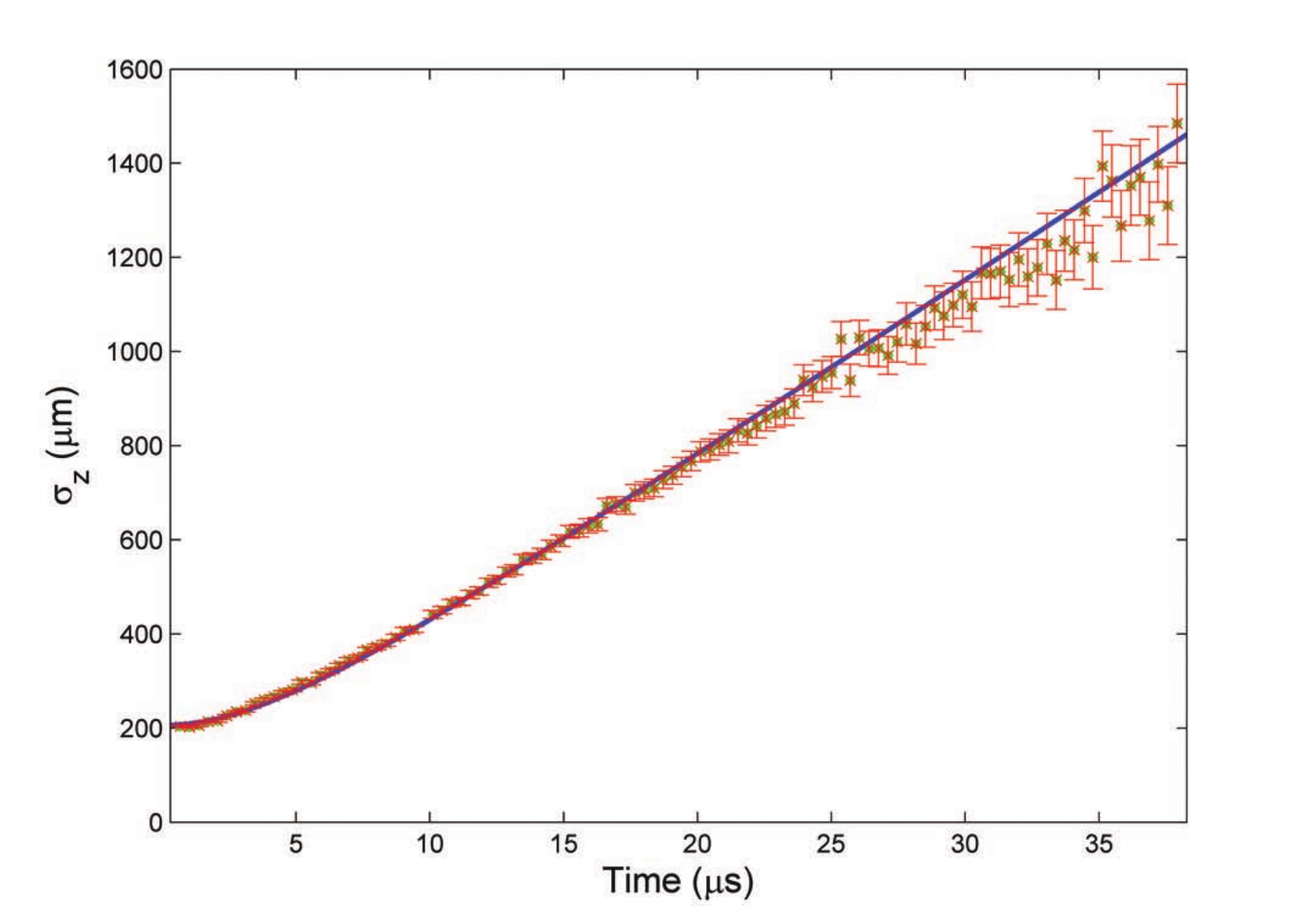}
    \end{center}
    \caption{Fitted plasma width as a function of centre arrival time at G$_2$. 
    Solid line is a fit of equation \ref{equ:sigma} to the experimental data, 
    yielding a fitted temperature $T_e + T_I=5$~K.}
    \label{fig:vlasovFit}
\end{figure}

Figure \ref{fig:stackplot} gives a sequence of electron signal traces recorded 
for an initial Rydberg state in the $nf$ series converging to $N^+=2$ for which 
we have selected the principal quantum number $n=50$ ($50f(2)$). Successive 
waveforms show electron signal amplitude as a function of time for increasing 
G$_2$ displacements. We see that the position of the late signal advances with 
the position of G$_2$, and that the arrival time of this feature corresponds in 
each case to the flight time of a neutral volume in the molecular beam from 
the excitation region to the perpendicular plane of G$_2$. 
This signal of electrons, extracted as the illuminated volume passes through 
G$_2$, broadens and decreases in amplitude with increasing flight time, which 
reflects the processes of expansion and decay of the plasma with time. Each 
trace fits well to a Gaussian function, and we use the parameters of such fits 
to extract the plasma arrival time, width and relative measure of the total number of electrons. Figure 
\ref{fig:vlasovFit} plots the fitted width along the axis of propagation (the 
short axis of the ellipsoid excitation volume), measured as a function of flight 
time. 

\section{Discussion}

\subsection{Expansion and electron temperature}

To approximately describe the long-time evolution of an expanding plasma we can refer to the Vlasov equations \cite{Pohl_thesis,Killian_Phys_rept}:

\begin{equation}
\frac{\partial f_\alpha}{\partial t}+\textbf{v}_\alpha \frac{\partial f_\alpha}{\partial \textbf{r}_\alpha} - m_\alpha^{-1} \frac{\partial f_\alpha}{\partial \textbf{v}_\alpha} q_\alpha \frac{\partial \phi(\textbf{r}_\alpha)}{\partial \textbf{r}_\alpha} =0 \label{vlas}
\end{equation}
Here, $\alpha =$ e,i denotes electrons and ions, respectively, $m_\alpha$ and $q_\alpha$ refer to the mass and charge of species $\alpha$, and $f_\alpha(\textbf{r}_\alpha,\textbf{v}_\alpha)$ represents the one-particle phase space probability density. The Vlasov equations for electrons and ions are coupled by the total mean-field potential $\phi$, which is determined by the Poisson equation

\begin{equation}
\Delta \phi = \frac{e}{\epsilon_0} (\rho_e - \rho_i),
\end{equation}
where $\rho_\alpha = \int f_\alpha d\textbf{v}_\alpha$. This problem does not yield a general, closed form analytical solution, and so we must employ several simplifying assumptions to approximate the evolution of our system.

As discussed previously, we take our initial Rydberg density to be that of a prolate ellipsoid.  We assume quasineutrality $\rho_e \approx \rho_i$.  We also assume that the heavy NO ions remain frozen on the timescale for plasma formation, and that the separable reservoir of electrons thermalizes instantly.  

Accordingly, we write the initial (t = 0) probability function for the ions as the calculated Rydberg phase space density, and express $f_e(\textbf{r},\textbf{v},t)$ for the electrons in terms of the quasistationary distribution:

\begin{equation}
f_e(\textbf{r},\textbf{v},t)= \rho_e(\textbf{r},t) * \phi_e^{(qs)}(\textbf{v},T_e(t)) \approx \rho_e(\textbf{r},t) * exp(-\frac{m_e v^2}{2 k_B T_e(t)})
\end{equation}
Using this expression in the Vlasov equation for electrons, together with the quasineutrality condition, yields an expression for the total mean-field potential

\begin{equation}
e \frac{\partial \phi}{\partial \textbf{r}}= k_B T_e \rho_e^{-1} \frac{\partial \rho_e}{\partial \textbf{r}} \approx k_B T_e \rho_i^{-1} \frac{\partial \rho_i}{\partial \textbf{r}}, \label{phi}
\end{equation}
which allows us to represent the ions by a Vlasov equation (\ref {vlasi}) that couples to the electrons via electron temperature and quasineutrality:

\begin{equation}
\frac{\partial f_i}{\partial t}+\textbf{v}_i \frac{\partial f_i}{\partial \textbf{r}_i} - \frac{k_B T_e(t)}{\rho ~m_i} \frac{\partial \rho}{\partial \textbf{r}_i} \frac{\partial f_i}{\partial \textbf{v}_i}  =0  
\label{vlasi}
\end{equation}
Dorozhkina and Semenov \cite{Semenov1,Semenov2} have derived an exact, self-similar solution to the Vlasov equations in the quasi-neutral approximation corresponding to the special case of a quadratic spatial dependence of the plasma electronic potential. Assuming Gaussian density distributions in $y,z$ and along $x$ in Eqs.(\ref{phi}), we find a solution to Eqs.(\ref{vlasi}):

\begin{equation}
f_i \propto exp (-\sum_k \frac{r_k^2}{2\sigma_k^2}) exp (-\sum_k \frac{m_i (v_k - \gamma_k r_k)^2}{2 k_B T_{i,k}}),
\end{equation}
where the subscripts $k=x,y,z$ represent the different cartesian coordinates, $\sigma_k$ is the rms-radius of the Gaussian spatial distribution, and $\gamma_k$ is the local hydrodynamic ion velocity. For expansion in the z-direction, over which we measure the plasma width, this solution yields a time-dependence for the radius $\sigma_z$, given by:

\begin{equation}
    \sigma_z(t)=\sigma_z(0)[1+t^2/\tau_z^{2}]^\frac{1}{2}\,
       \label{equ:sigma}
\end{equation}

\begin{equation}
    \tau_z^{2}=\frac{m_{i}\sigma_z(0)^2}{k_{B}[T_{e}(0)+T_{i}(0)]}\,
\end{equation}

We fit Gaussian radii, measured as a function of flight time, to Eq.~(\ref{equ:sigma}).  We extrapolate to determine the initial width of the plasma along the z-axis, $\sigma_z(0)$, and, from the rate of plasma expansion, $\sigma_z(t)$, we obtain $\tau_z$, which yields the sum of initial electron and ion temperatures.  We note that the time-dependent width displayed in Figure \ref{fig:vlasovFit} 
shows evidence of the acceleration characteristic of an electron-charge-driven 
ambipolar expansion. A Vlasov fit to this $\sigma(t)$ returns an initial 
electron temperature of 5~K. 

At the peak electron density of our plasma, the Wigner-Seitz radius, $a$, is 500~nm, 
and such a $T_e$ would indicate significant correlation ($\Gamma_{e}=7$, where 
$\Gamma_{e}=q^{2} /  4\pi a\epsilon_{0}k_{B}T_e$). The Vlasov equations do not 
extend with accuracy to describe the expansion of a charged-particle systems under such 
conditions. At high correlation, the positive field created by the cations contained in 
the volume of the plasma lowers the potential energy of free electrons. The depth of this well decreases with expansion, and this elevation of potential energy with expansion acts as a force that 
suppresses electron pressure \cite{Murillo}. Thus, for the purposes of the following 
discussion, we can assume that the initial temperature we estimate from the rate of plasma 
expansion represents a lower limit. 

\subsection{Kinetic model for the evolution to plasma}

Double-resonant excitation transfers the order of $3\times 10^9$ NO molecules in a volume under 1 mm$^3$ to a selected Rydberg state (e.g. $n = 50$, some 40 cm$^{-1}$ below the ionization threshold).  Prompt electron emission signifies a rapid evolution to plasma.  

The ions expand, as described above, with a radial velocity that characteristically accelerates in response to the hydrodynamic force of the expanding electrons.  The observed rate at which this molecular beam plasma expands significantly lags the expansion rates of plasmas produced from atoms cooled in a MOT, which suggests a lower electron temperature.  The question arises of whether we can account for such conditions with the context of conventional kinetic models for ultracold plasma evolution.  

To explore this question, we have adapted a conventional model for ion-electron-atom equilibration in ultracold atomic plasmas \cite {PPR2003,PPR2004,Pohl_atom} to this molecular case:  Penning ionization in the Rydberg gas precedes electron evaporation followed by an electron impact avalanche to a quasi-steady state consisting of molecular ions, electrons and excited NO molecules distributed over a broad interval of Rydberg states.  Sections to follow detail the approach we use to model these initial steps and present results describing the relaxation of a model atomic system of fixed total density.  Then finally, in the context of these computed results, we consider the possibility of new physics owing to the effects of Rydberg predissociation and electron-ion dissociative recombination at work in an expanding molecular plasma of non-uniform density.

\subsubsection{Penning ionization}

Our experiment creates a gas of Rydberg molecules with a thousand-fold higher density than the gas of excited atoms typically produced in a MOT.  Comparatively strong dipole-dipole forces drive the interactions between these species \cite{Pillet2008}.  An NO* molecule, excited to a principal quantum number, $n_0=50$, has a diameter of 260 nm.  In the core of the illuminated volume, the average distance between such molecules is about 500 nm.  Classical simulations suggest that Penning ionization will occur within 1000  Rydberg periods for pairs of NO* molecules positioned within two orbital diameter of each other  \cite{Robicheaux2005p}.  To find the fraction of such molecules in a Rydberg gas of density $\rho$, we refer to the Erlang distribution of nearest neighbour distance, $r$ \cite{Erlang}.  

 \begin{equation}
    d\xi_{\rho}(r)=4 \pi \rho r^2 e^{-\frac{4\pi}{3}\rho r^3}dr
       \label{equ:Erlang}
\end{equation}

To estimate a total number of Penning electrons produced over the illuminated volume, we consider the fraction of molecules within twice the classical diameter of the selected Rydberg state, and combine Penning fractions for regions of approximately constant $\rho$.  Total integrated Penning fractions calculated in this way range from 10 percent for $n_0=30$ to 50 percent for $n_0=80$.  

Penning ionization yields a distribution of companion NO$^*$ molecules in Rydberg states of lower principal quantum number.  By energy conservation, the maximum principal quantum number cannot exceed $n=n_0/ \sqrt{2}$.  The distribution of final states over $n$ directly reflects the distribution of Penning electron kinetic energies, and thus the temperature of the Penning electrons.  As discussed by Robicheaux \cite{Robicheaux2005p}, this long-range Coulomb interaction favours the production of threshold electrons, and in a classical Monte Carlo model for Penning ionization, he finds that $n$ declines from the maximum as $n^5$.  Fitting a Maxwell-Boltzmann distribution to the scaled distribution of calculated electron kinetic energies in \cite{Robicheaux2005p}, we observe that the initial Rydberg binding energy predicts the Penning electron temperature, $T'_e$,  as $3 k_B T'_e \approx  \mathcal{R}/ n_0^2$.  On this basis, we estimate initial electron temperatures following Penning ionization that range from 58~K for $n_0=30$ to 8 K for $n_0=80$.

\subsubsection{Electron evaporation}

The first Penning electrons escape until the space charge created by the ions left behind rises sufficiently to trap the electrons that remain.  Following the work of Comparat et al. \cite{Comparat_star}, we can estimate the excess charge $(N_i-N_e)$ required to trap electrons of energy $k_BT_e$:

\begin{equation}
N_i-N_e=\frac{3}{2}k_BT_e \sigma \frac{4\pi\epsilon_0}{e^2}\sqrt{\frac{\pi}{2}}
\label{equ:ionfrac}
\end{equation} 

\noindent The rate equation model described below accounts for this loss of energetic electrons and the development of a trapping potential.  Using Eq. \ref{equ:ionfrac}, we calculate that the electron temperature, as it evolves from Penning ionization through the avalanche to steady state, supports the evaporation of 0.1 percent of the electrons, reducing the electron temperature by less than one percent.    

\subsubsection{Electron impact}

Trapped electrons collide with Rydberg molecules and NO$^+$ ions. This leads to electron impact redistribution:
\begin{eqnarray}
{\rm NO}^* + e^-  & \rates{ k_{n_i,n_f}}{ k_{n_f,n_i}} &{\rm NO}^{**}  + e^-
\label{equ:kij}
\end{eqnarray}
\noindent together with electron impact ionization and three body recombination (TBR):  
\begin{eqnarray}
{\rm NO^*}  + e^-& \rates{k_{ion}}{k_{tbr}} &{\rm NO}^+ + e^-  + e^- 
\label{equ:kion}
\end{eqnarray}

The diatomic nature of NO$^+$ and NO$^*$ introduces two additional processes unique to molecular plasmas:  Dissociative recombination (DR):
\begin{eqnarray}
{\rm NO^{+}} + e^-
 & \displaystyle \mathop{\longrightarrow}^\textrm{\scriptsize{k$_{DR}$}}
 & {\rm N} + {\rm O}
\label{equ:DR}
\end{eqnarray}

\noindent and the state-detailed unimolecular predissociation (PD) of Rydberg molecules:
\begin{eqnarray}
{\rm NO^{*}} 
 & \displaystyle \mathop{\longrightarrow}^\textrm{\scriptsize{k$_{PD(n)}$}}
 & {\rm N} + {\rm O}
\label{equ:PD}
\end{eqnarray}

Processes (\ref{equ:kij}) and (\ref{equ:kion}) govern the evolution to plasma and the formation of a steady-state distribution of ions, electrons and Rydberg molecules.  Dissociative recombination controls the long-time dissipation of the plasma.  Experiments and theory provide a reliable estimate of $k_{{DR}}$ as a function of $T_e$, from which we can conclude that DR proceeds too slowly to affect the short-time dynamics of the plasma under our conditions.  So, for the present purposes, we neglect dissociative recombination.  For now, we also neglect Rydberg state predissociation, with the idea of returning to this question later.  

\begin{figure}[ht]
    \begin{center}
      \includegraphics[height=4in]{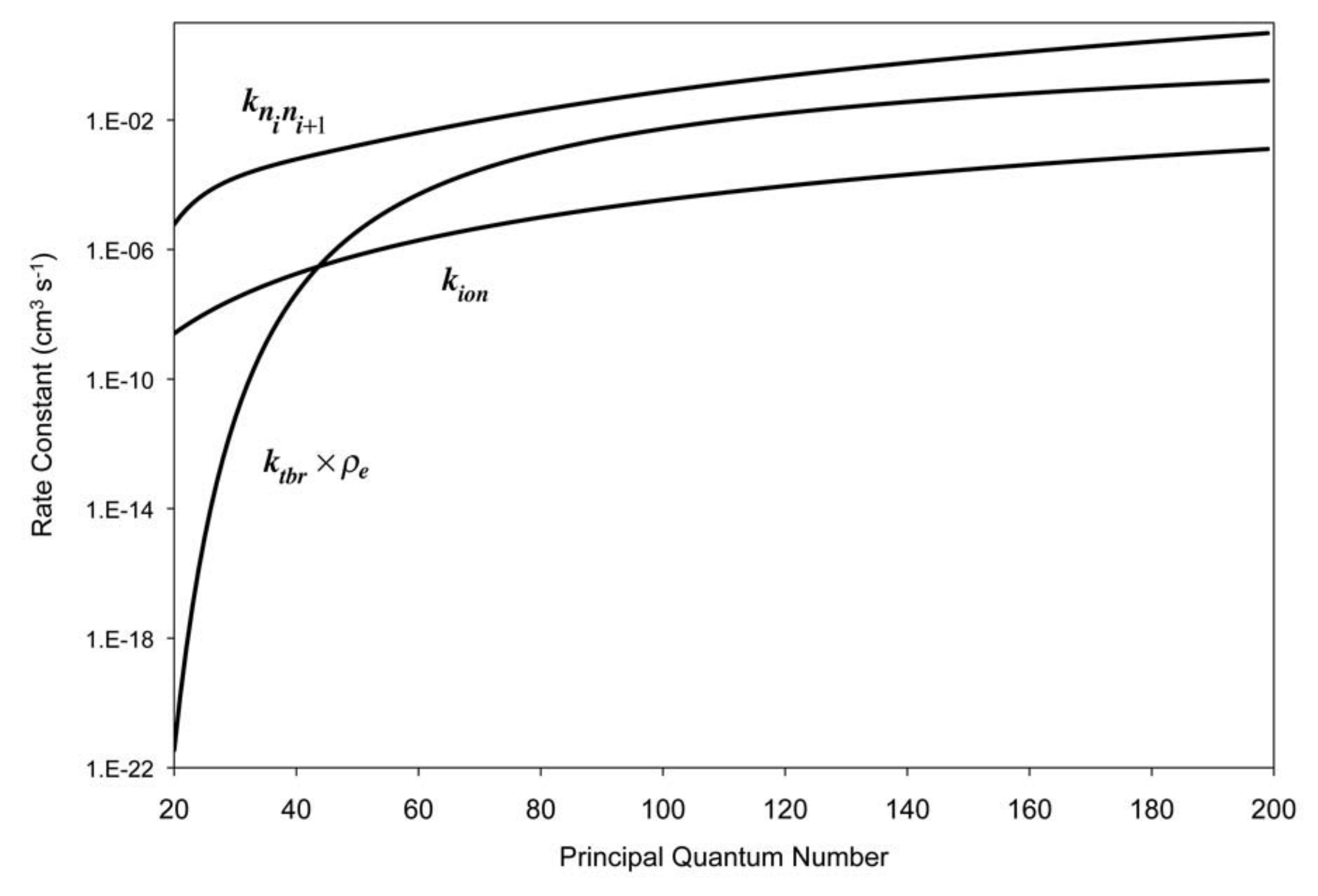}
    \end{center}
    \caption{Rate constants for:  Electron-impact collisional excitation of Rydberg states by $\Delta n = 1$, $k_{n_i n_{i+1}}$; Three-body recombination to form Rydberg state, $n$, $k_{tbr}$ (multiplied by an electron density of $10^{10}$ cm$^{-1}$); Electron impact ionization of Rydberg state, $n$, $k_{ion}$, for an electron temperature, $T_e$ = 10 K}
    \label{fig:rateconstants.pdf}
\end{figure}

With these simplifications, the evolution to plasma parallels entirely the process characterized for atomic systems in MOTs.  We write coupled differential equations for this with reference to theoretical rate constants refined recently by comparison with Monte Carlo simulations.  Theoretical rate constants derived in this way depend on $T_e$ and relevant Rydberg principal quantum numbers.  Figure (\ref{fig:rateconstants.pdf}) plots representative values of $k_{ ion}$, $k_{tbr}$ and $k_{n,n+1}$, in each case as a function of $n$.

We use these rate constants to simulate the short-time dynamics of plasma formation.  By numerically solving a set of (N+1) coupled differential rate equations, we describe the time evolution of population density between a distribution of free ions and electrons and N Rydberg levels ranging in principal quantum number, $n$, from 25 to 125. The model assumes a uniform, conserved density of particles and energy distributed over free electrons and bound Rydberg molecules.  

\begin{figure}[ht]
    \begin{center}
      \includegraphics[height=4in]{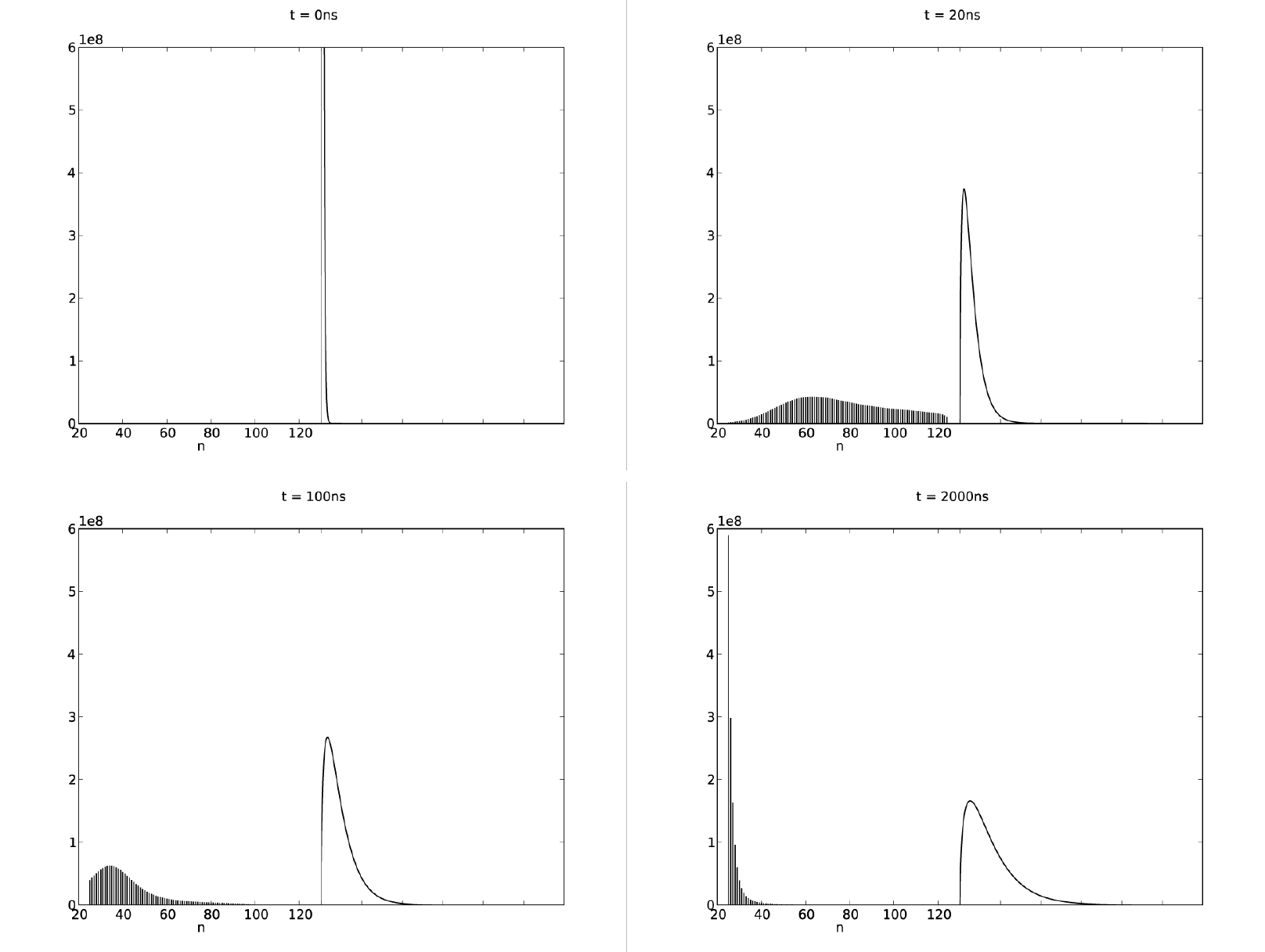}
    \end{center}
    \caption{Simulation results showing Rydberg level populations (vertical bars) and distribution of free-electron energy for specified times following the formation of a plasma with a uniform ion density of $10^{10}$ cm$^{-3}$ and initial $T_e$ = 1 K.  Electron density and temperature for times following $t=0$:  $t=20$ ns, $7.5 \times 10^{9}$ cm$^{-3}$ $T_e$ = 9.7 K; $t=100$ ns, $8.4 \times 10^{9}$ cm$^{-3}$ $T_e$ = 15 K; $t=2$ $\mu$s, $8.6 \times 10^{9}$ cm$^{-3}$ $T_e$ = 25 K.}
    \label{fig: Frames.pdf}
\end{figure}

To initialize this calculation, we either define a $t=0$ distribution of free electrons, created from a specified Rydberg state population by Penning ionization, or choose a density and temperature of free electrons and ions.  When applied to model an atomic plasma, prepared by threshold photoionization under MOT conditions, this calculation conforms well with published experimental results \cite{Laha}.  Figure \ref{fig: Frames.pdf} shows a sequence of frames that demonstrate the relaxation of a hypothetical plasma with an initial electron density $10^{10}$ cm$^{-3}$ and $T_e$ = 1 K.  Rydberg states form by three-body recombination and undergo superelastic collisions heating electrons to a steady-state temperature of 25 K, as seen in MOT experiments.  

Less is known experimentally about the steady-state conditions of ultracold plasmas ignited by Rydberg-Rydberg collisions.  We find that model calculations, using the same rate coefficient basis, yield nearly the same steady state for a plasma formed by the evolution of a gas of $n_0=50$ Rydberg states from an initial step of Penning ionization to an electron impact avalanche of energy-redistributing, ionizing and ion-electron recombining collisions.  For a MOT-like density of $10^{10}$ cm$^{-3}$, such simulations predict evolution over a time interval of 2 $\mu$s to a steady-state electron density of $6.2 \times10^{10}$ cm$^{-3}$ and temperature, $T_e$ = 22 K.  

Our experimental ion/electron-plus-Rydberg density exceeds that typically found under MOT conditions by three orders of magnitude.  Moreover, molecular degrees of freedom open channels for the loss of particles and ion-electron binding energy.  Our data suggest plasma steady states of much lower electron temperature.  In the context of a rate equation model, can these conditions particular to a molecular beam plasma explain this apparent shift in steady-state?  

Increasing density alone shortens the timescale for relaxation and raises the steady-state electron temperature.  A larger Penning fraction, together with faster electron impact energy transfer and ionization accelerates the avalanche, and proportionately faster three-body recombination transfers more Rydberg binding energy to a proportionally smaller reservoir of free electrons.  In model calculations, a gas of $n_0=50$ Rydberg states with a density of $2\times 10^{12}$ cm$^{-3}$ relaxes in 5 ns to a steady-state electron density of $1\times 10^{12}$ cm$^{-3}$ and temperature, $T_e$ = 37 K.  

What then of Rydberg predissociation?  Most of the electron heating associated with three-body recombination arises from super-elastic electron-impact collisions that deactivate Rydberg states of relatively low principal quantum number.  Could predissociation diminish this heating by removing Rydberg states formed by three-body recombination before they relax to lower $n$?  

For model systems using conventional rate constant formulations for $k_{n_i,n_f}$, $k_{ ion}$ and $k_{tbr}$, the answer is decidedly no.  The electron-Rydberg and electron-ion rate processes associated with the evolution to steady-state maintain a Saha equilibrium that rapidly refills any level population lost to predissociation.  Under these conditions, predissociation always produces net heating.  In a separate study of long-time plasma decay kinetics, we have found experimental plasma lifetimes that accord with rate coefficients observed for conventional two-body dissociative recombination of NO$^+$, with little discernible contribution from the unimolecular dissociation of Rydberg states \cite{Rennick2011}.  In model calculations, if we accelerate predissociation to a degree sufficient to affect Rydberg level-population distributions, we rapidly lose all of the NO$^+$ ions and electrons in the conversion to neutral N and O atoms.  

A model with the capacity to produce the low-$T_e$ free electron distribution that we see experimentally requires several particular characteristics.  Collisional ionization must occur with an efficiency comparable to collisional energy redistribution.  The model must suppress three-body electron-ion recombination.  And, some early-stage process must intercept the super-elastic electron-impact cascade to lower-$n$ Rydberg states.  

The enhancement of collisional ionization and suppression of three-body recombination have both been suggested as consequences of strong coupling \cite{Schlanges,Pattard2006}.  As noted above, it seems unlikely that spontaneous predissociation can keep pace with the rate at which collisions with electrons redistribute the energy of Rydberg molecules.  However, with 8 eV of internal energy NO$^*$ presents a dense manifold of underlying valence states \cite{NO_valance}.  Strong perturbations arising in electron-molecule collisions could go far to relax configurational barriers to Rydberg-valance coupling and consequent intramolecular relaxation \cite{Jungen}.  Excited states populated in this way would be lost to the Rydberg cascade.  

The plasma formed at high density in a molecular beam clearly challenges conventional understanding.  We are presently at work in an effort to develop a comprehensive model, in which we incorporate effects of strong coupling and multichannel collisional energy transfer, in a system that evolves from a molecular Rydberg gas such as this to a plasma under conditions of both spatial and time varying density.   

\section{Acknowledgements}

We are pleased to acknowledge helpful discussions with T. Pohl and J. M. Rost.  This work was supported by the Natural Sciences and Engineering Research Council of Canada (NSERC), the Canada Foundation for Innovation (CFI) and the British Columbia Knowledge Development Fund (BCKDF).

\section{References}

\bibliography{ref}

\enddocument